\documentstyle[11pt,aaspp4]{article}
\input epsf
\lefthead{SANDQUIST ET AL.}
\righthead{DOUBLE CORE EVOLUTION. X.}
\def\n{\noindent}
\def\msun{\mbox{M$_{\odot}$}}
\def\rsun{\mbox{R$_{\odot}$}}
\def\lsun{\mbox{L$_{\odot}$}}
\def\lapprox{\;\rlap{\lower 2.5pt                       
             \hbox{$\sim$}}\raise 1.5pt\hbox{$<$}\;}
\def\mpy{M$_{\odot}$ yr$^{-1}$}
\begin{document}
\title{DOUBLE CORE EVOLUTION X. THROUGH THE ENVELOPE EJECTION PHASE}
\author{Eric L. Sandquist\altaffilmark{1}, Ronald E. Taam\altaffilmark{1},
Xingming Chen\altaffilmark{2}, Peter Bodenheimer\altaffilmark{2}, and
Andreas Burkert\altaffilmark{3}}

\n \altaffilmark{1}{Department of Physics \& Astronomy, Northwestern 
University, Evanston, IL 60208; erics@apollo.astro.nwu.edu, 
taam@apollo.astro.nwu.edu}

\n \altaffilmark{2}{UCO/Lick Observatory, Board of Studies in Astronomy
and Astrophysics, University of California, Santa Cruz, CA, 95064;
chen@ucolick.org, peter@ucolick.org}

\n \altaffilmark{3}{Max Planck Institut for Astronomy, Koenigstuhl 17, 
D-69117 Heidelberg, Germany; burkert@mpia-hd.mpg.de}

\dates

\begin{abstract}
The evolution of binary systems consisting of an asymptotic giant
branch star of mass equal to $3 \msun$ or $5 \msun$, and a main
sequence star of mass equal to $0.4 \msun$ or $0.6 \msun$ with orbital
periods $\gtrsim 200$ days has been followed from the onset through
the late stages of the common envelope phase.  Using a nested grid
technique, the three-dimensional hydrodynamical simulations of an
asymptotic giant branch star with radii $\sim 1$ A.U. indicate that a
significant fraction of the envelope gas is unbound ($\sim 31$\% and
23\% for binaries of $3 \msun$ and $0.4 \msun$, and $5 \msun$ and $0.6
\msun$ respectively) by the ends of the simulations, and that the
efficiency of the mass ejection process $\sim 40$\%.  During an
intermediate phase, a differentially rotating structure resembling a
thick disk surrounds the remnant binary briefly before energy input
from the orbits of the companion and remnant core drive the mass away.
While the original volume of the giant is virtually evacuated in the
late stages, most of the envelope gas remains marginally bound on the
grid. At the ends of our simulations, when the orbital decay timescale
exceeds about 5 years, the giant core and companion orbit one another
with a period of $\sim 1$ day (2.4 days for a binary involving a more
evolved giant), although this is an upper limit to the
final orbital period. For a binary of $5 \msun$ and $0.4 \msun$, the
common envelope may not be completely ejected.  The results are not
found to be sensitive to the degree to which the initial binary system
departs from the synchronous state.
\end{abstract}


\section{Introduction}\label{intro}
There are many classes of binary systems which cannot be understood
without a phase of significant mass and angular momentum loss.  Among
those classes are interacting binary systems containing white dwarf or
neutron star components.  Common envelope evolution has been
identified as a means by which the interactions between the two
progenitor components lead to the ejection of a large fraction of the
mass of the system and to the shrinkage of the binary orbit (see
Paczynski 1976). Provided that the common envelope is successfully
ejected, a long-period system ($P \sim$ years) can be transformed into
a short-period system ($P \sim$ days) consisting of the remnant core
of the red giant and its companion.

A binary system can evolve into the common envelope stage whenever the
red giant progenitor of the compact component enters a phase where it
is not corotating with respect to the orbit at the onset of or
during mass transfer to its companion.  This may occur when mass
transfer from the more massive star to its companion is unstable, as
occurs when the Roche lobe contracts faster than its radius,
especially in the case of a giant with a deep convective envelope (see
Paczynski \& Sienkiewicz 1972; Webbink 1979).  Another evolutionary
path leading to the common envelope stage involves a tidal instability
(Darwin 1879; Counselman 1973; Kopal 1978; Lai, Rasio, \& Shapiro
1993, 1994), which occurs when a system reaches a minimum value of the
total angular momentum for a synchronized close binary system.
Evolution to this stage inevitably leads to the infall of one
component into its more massive companion.

As a result of the development of powerful numerical algorithms and
the advancement of computer technology, progress has been made in
modeling common envelope interactions from the slow initial phase to the
rapid phase of orbital shrinkage.  Many of the simplifying
approximations of the past regarding the input physics, as well as the
dimensionality of the problem, have been relaxed.  In particular, the
multidimensional studies of Bodenheimer \& Taam (1984), Livio \&
Soker (1988), and Taam \& Bodenheimer (1989) revealed that matter is
preferentially ejected at higher than the escape speed along the
equatorial plane of the binary system.  This theoretical result
received observational support from the elliptical or butterfly-shaped
appearance of planetary nebulae with binary nuclei (Bond \& Livio
1990).  That is, the morphology of such nebulae are consistent with
the density contrast between the pole and equator seen in common
envelope simulations used in interacting winds models (Kwok 1982;
Soker \& Livio 1989; Frank et al. 1993; Mellema \& Frank 1995; Livio
1995). This nonspherical ejection follows from the angular
distribution of the energy and angular momentum deposited in the
common envelope by the orbital motion of the two components.

Population synthesis calculations by a number of investigators (see,
for example, Tutukov and Yungelson 1979; Iben \& Tutukov 1984; van
den Heuvel 1987; de Kool 1992; Iben \& Livio 1993; Yungelson,
Tutukov, \& Livio 1993; Han, Podsiadlowski, \& Eggleton 1995; Kalogera
\& Webbink 1996; de Kool 1996) have used approximations to the common
envelope phase to predict the properties of systems which survive.
Such calculations make use of energetic arguments, and the results are
sensitive to the adopted value of the efficiency of the common
envelope phase (taken to be equal to the ratio of the binding energy
of the ejected mass to the energy lost from the binary orbit).
Although the results of the early numerical studies of Taam,
Bodenheimer, \& Ostriker (1978), Meyer \& Meyer-Hofmeister (1979),
Delgado (1980), Bodenheimer \& Taam (1984), Livio \& Soker (1984,
1988), and Taam \& Bodenheimer (1989, 1991) were instructive, the
important details of the outcome of the common envelope phase were
still unclear, since understanding of the mass ejection process was
affected by the approximations (such as using smaller numbers of
spatial dimensions, or imposed descriptions for the orbital energy
dissipation) needed to make the problem tractable.  Thus, the
production of short-period binary systems containing compact objects
remained to be clarified (even if sufficient energy was lost from the
orbit to unbind the common envelope) since the results of detailed
calculations have not been taken into account in the population
synthesis studies in determining whether the two stars would stop
spiraling toward each other (see Taam \& Bodenheimer 1991).
 
In this paper, we investigate the details of the hydrodynamics of the
common envelope phase in the vicinity of the double core at high
spatial resolution in three dimensions in order to understand the
termination of the spiral-in phase.  Previous three-dimensional
studies (de Kool 1987; Terman et al. 1994, 1995; Terman \& Taam 1996;
Rasio \& Livio 1996) have relied on the smoothed particle
hydrodynamics technique (SPH; e.g., Lucy 1977; Gingold \& Monaghan
1977; Monaghan 1985, 1992).  The numerical results of such studies
indicated that mass is ejected from the common envelope, although the
poor spatial resolution of the simulations made it unclear whether the
companion would spiral into the red giant core.  Although the SPH
technique can model the global features of the evolution during
earlier stages, our experience indicates that the numerical resolution
is not entirely adequate to model the region of steep density
gradients above the red giant core during the terminal phase of
evolution.  In addition, the large number of time steps required to
follow the ejection of a significant fraction of the envelope using
higher resolution is computationally prohibitive.

To adequately resolve the region close around the double core, we use
a finite differencing numerical technique. In contrast to the study of
Livio \& Soker (1988), we use an Eulerian nested grid technique
similar to that used in Yorke et al. (1995), but extended to three
dimensions.  We report on the results of simulations of the common
envelope phase for binaries consisting of an intermediate-mass
asymptotic giant branch star of either 3 or 5 $\msun$ with a main
sequence companion of 0.4 or 0.6 $\msun$.  In the following section we
describe the numerical methods used, the assumptions underlying the
calculations, and the construction of the initial models.  The
detailed evolution of a binary from the initial spiral-in phase
through the late phase of envelope ejection is presented in \S 3,
as are comparisons with our other simulations, and a discussion of
physics affecting the system beyond the scope of our simulations.
The implications of the simulations are discussed in the final
section.

\bigskip

\section{Numerical Methods}

For all of the simulations in this paper, a 3-D grid-based
hydrodynamics code is used.  A nested grid technique is employed to
provide higher spatial resolution in the inner regions of the common
envelope, with dynamical coupling between grids based on the
technique of Berger \& Oliger (1984) and Berger \& Colella (1989), as
described in Yorke, Bodenheimer, \& Laughlin (1993) and Ruffert
(1993).  We have modified a version of the code used by Burkert \&
Bodenheimer (1993), which they used to follow the isothermal
gravitational collapse of a protostar. In our simulations, the main
grid has $64 \times 64 \times 64$ cubical zones, while the nested
subgrids have $64 \times 64 \times 32$ cubical zones, where the short
dimension is perpendicular to the orbital plane. Two subgrids were
centered within the main grid, and were kept motionless with respect
to it. The first subgrid was a factor of 4 smaller in the $x$ and $y$
directions, and the second subgrid was a factor of 2 smaller than the
first, so that the zone size on the finest grid was $1.56 \times
10^{11}$ cm in all but one of the simulations (for which it was twice
as large). The total mass, energy, and angular momentum of the gas lost
from the main grid were followed.

In order to simulate realistic interactions between a giant star and a
main sequence companion, the code also follows the motions of two
collisionless particles --- one for the core of the red giant star,
and one for the companion. Gravitational interactions between the
collisionless particles and the gas were smoothed in a manner
suggested by Ruffert (1993):
\[ \Phi_{PG} = \frac{-G M_{P}}{\sqrt{r^{2} + \epsilon^{2} \delta^{2}
\exp(-(r/\epsilon \delta)^2)}} ,\]
where $\delta$ is the width of a zone on the innermost subgrid, and
$\epsilon = 1.5$. Here $r$ is the distance between the gas and the particle
of mass, $M_P$. The gravitational potential of the gas was computed
separately on the main grid and subgrids via fast Fourier
transforms. Boundary values for the subgrid computations were taken
from the smallest grid that encompassed it.

Timesteps for each grid were chosen according to a
Courant-Friedrichs-Lewy condition on that grid:
\[ \Delta t = C \delta_{i} / \max(\vert u_{x} \vert + \vert u_{y} \vert +
\vert u_{z} \vert + c_{s}) ,\] where $C = 0.5$ is the Courant factor,
and $c_{s}$ is the sound speed. On the innermost subgrid, the timestep
can be reduced to
\[ \Delta t = C \delta / \max(\vert v_{1} \vert, \vert v_{2} \vert)\]
if this is smaller than the timestep derived from the gas, where
$v_{1}$ and $v_{2}$ are the speeds of the two collisionless particles.
Changes to the Courant number were made as tests, but these had no
effect on the results.

Core particles were stepped in position using a four-point Runge-Kutta
integration scheme each time the gas distribution was recalculated on
the innermost subgrid. When the cores are at their closest approach,
the orbits only pass through a handful of zones, so that with the
Courant-type timestep, the final orbit would be poorly integrated.  To
improve on this, the gas timestep was increasingly subdivided to ensure
that the core positions are computed at least 25 times during each
orbit.

Because the original version of the code was intended for use in
computing isothermal collapse, the code was modified to explicitly treat 
the internal energy of the gas. The equation of state
was chosen to be a combination of ideal gas and radiation
pressure (the regions where the gas is affected by electron degeneracy
are sufficiently compact that they can be assumed to be restricted to the
point masses). Internal energy density was advected in exactly
the same manner as mass density, and compressional heating of
the gas was also included.  Shocks in the gas were handled with an
artificial viscosity of the type introduced by von Neumann and
Richtmeyer (1950) with a constant shock smoothing length of
$1.18 \delta_{i}$, where $\delta_{i}$ is the zone size in
grid or subgrid $i$. Advection of density, internal energy, and
momentum across zone boundaries was handled with a monotonic transport
formalism (van Leer 1977), which provides second-order spatial
accuracy, and quasi-second order accuracy in time.

A number of additional tests were conducted in order to check the
accuracy of the code for common envelope simulations. First, we
conducted a standard shock tube run (Sod 1978), verifying that the
numerical results matched the analytical solution for adiabatic flow.
The results were substantially similar to those of Burkert \&
Bodenheimer (1993).  We verified that the shocks were not disturbed by
the presence or absence of subgrids in the calculation.  Secondly,
several trial common-envelope runs were conducted to check the
accuracy of various aspects of the code. The most important effects
involved the smoothing length used in gravitational interactions
between core particles and the gas. The value of the smoothing length
$\epsilon \delta$ adopted is a minimum acceptable value --- lower
values resulted in unacceptable non-conservation of total energy.
Because of the importance of understanding the orbital evolution of
the core particles after the majority of the gas has been forced away
during the infall of the companion, the smoothing length was reduced
as much as possible.  The same gravitational smoothing length was used
in all levels of each simulation since it was found that significant
non-conservation of total energy occurred if the point mass came near
the edge of a subgrid during simulations in which the smoothing length
was made equal to $\epsilon$ times the size of a zone in the grid
which the particle was situated.

The degree of physical precision of our calculations depends on the
resolution of the gas in the regions of strongest interaction (the
number of subgrids used) and on the accuracy of the force calculation
near the core particles (the core-gas smoothing length). The degree to
which energy was conserved depended on the maximum strength of the
core gravitational fields (which was determined by the smoothing
length). A similar effect was found by Ruffert (1993), in that the
density and temperature structure near the core was sensitive to the
structure of the core potentials. After a number of tests, it was
found that for a given smoothing length, fewer subgrids resulted in
slightly better conservation of energy. This motivated our use of only
two subgrids. We will return to this topic in \S~\ref{numerical},
where we discuss the effects of resolution on the final orbital
separation.
\bigskip

\subsection{Initial Models}

The initial gas distribution in the envelope of the red giant was
derived from one-dimensional stellar models obtained from
the code developed by Eggleton (1971, 1972). Interpolation
within a stellar model in the appropriate evolutionary state was used
to obtain densities and internal energies for the zones. A diffuse
background gas filled the remainder of the grid, and was given a
temperature to bring it into pressure equilibrium with the surface of
the star. The mass of the core particle for the giant was chosen to
bring the total giant mass to that of the input stellar model.  The
orbit of the binary was placed in the xy-plane. We also
forced the gas to be symmetric about this plane, so that at all times
the core particles remained very close to z$=0$.

In order to determine the sensitivities of the outcome of the common
envelope evolution, we have varied three parameters: the companion
mass, the red giant mass, and the fraction of synchronous rotation
given to the envelope of the giant.  Our baseline simulation, against
which we will compare all of our runs, involved a $3 M_{\odot}$
red giant star (with a $0.7 M_{\odot}$ core, and a total radius of
$1.39 \times 10^{13}$ cm), and a $0.4 M_{\odot}$ companion particle.
The companion was initially placed in a circular orbit at a distance of
$2 \times 10^{13}$ cm, with the giant in synchronous rotation with the
companion. For the other simulations, these parameters were
varied, and the alternate values used were: a $0.6
M_{\odot}$ companion; a $5 M_{\odot}$ giant star (with the same radius
as the $3 M_{\odot}$ giant, but with a $1 M_{\odot}$ core); and a
non-rotating giant. One additional simulation involving a more
evolved $5 M_{\odot}$ giant with a radius of $2.46 \times 10^{13}$ cm
was run with a companion of $0.6 M_{\odot}$ orbiting at a distance of
$3.7 \times 10^{13}$ cm.

We have not started our baseline simulation in a relaxed-binary
configuration as did Rasio \& Livio (1996), but we believe that the
difference in the initial conditions is a minor consideration. In the
Rasio \& Livio's (1996) case, matter was barely
overflowing the giant's Roche lobe whereas in our initial
configuration the mass transfer developed quickly.  This difference in
initial conditions results in a sharper angle of impact for the
companion star onto the giant's surface in our simulation. In
addition, the efficiency of mass ejection from the binary system is
somewhat underestimated in the present study since the energy transfer
from the orbit to the giant envelope to bring it to a Roche-lobe
filling configuration is already incorporated into the envelope in the
initial state described by Rasio \& Livio (1996).  However, this
difference is not important for the final outcome since this aspect of
the evolution affects only the near-surface gas, which is the least
bound material. Hence, the energy difference is rather small in that
sense.  Most importantly, the general agreement in the final state of
our simulations between the synchronously rotating and
asynchronously rotating cases argues that the effects of the initial
conditions are minor. In particular, the evolution and timescales for
the rapid dynamical phase and mass ejection phase are unaffected by
these details of the initial configuration.

\section{Results}

In this section we report on the numerical results of five
simulations. The initial parameters and results for each
sequence are summarized in Tables~1 and 2 respectively. For
convenience, the results are presented in detail for the baseline
simulation (sequence 1) followed by descriptions of the
differences that exist among the simulations. For sequence 1, we
consider the common envelope evolution of a binary consisting of a $3
\msun$ asymptotic giant branch (AGB) star with a $0.4 \msun$ main
sequence companion. The two components are orbiting about their
common center of mass in circular motion with a period of 1.2 years.

\subsection{Spiral-In Phase}

\begin{figure}
\hspace*{-0.5in}
\epsffile{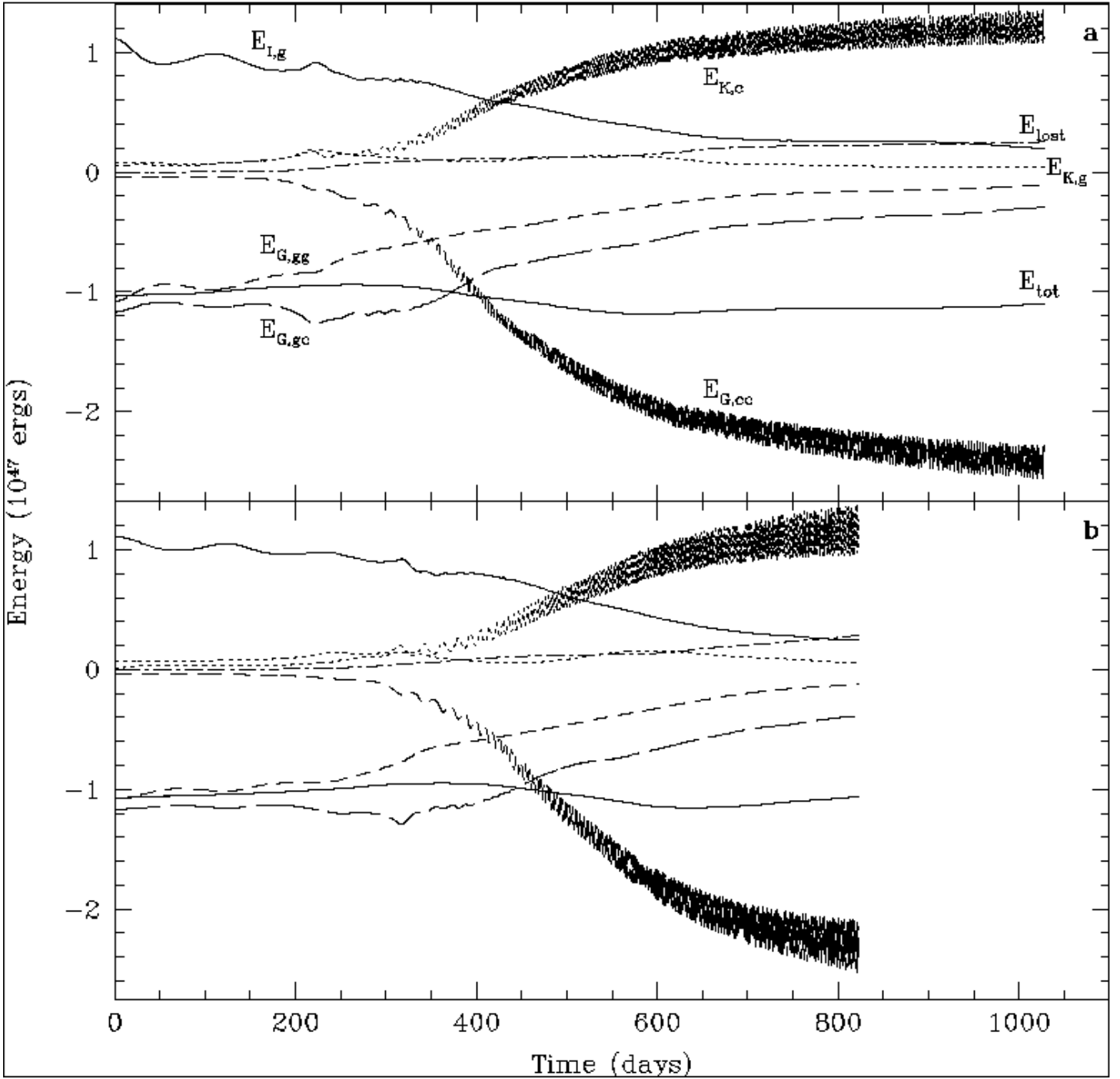}
\figcaption{The contributions to the energy
for a) simulation 1 and b) simulation 2 (no initial
rotation of the gas). The curves from top to bottom at 700 days are:
kinetic energy of the point masses ($E_{K,c}$), internal energy of the
gas ($E_{I,g}$), energy lost from the grid ($E_{lost}$), kinetic
energy of the gas ($E_{K,g}$), potential energy from gas-gas
interactions ($E_{G,gg}$), and from gas-point mass interactions
($E_{G,gc}$), total energy, and potential energy from core-companion
(point mass) interactions ($E_{G,cc}$). \label{erg1}}
\end{figure}

\begin{figure}
\hspace*{-0.5in}
\epsffile{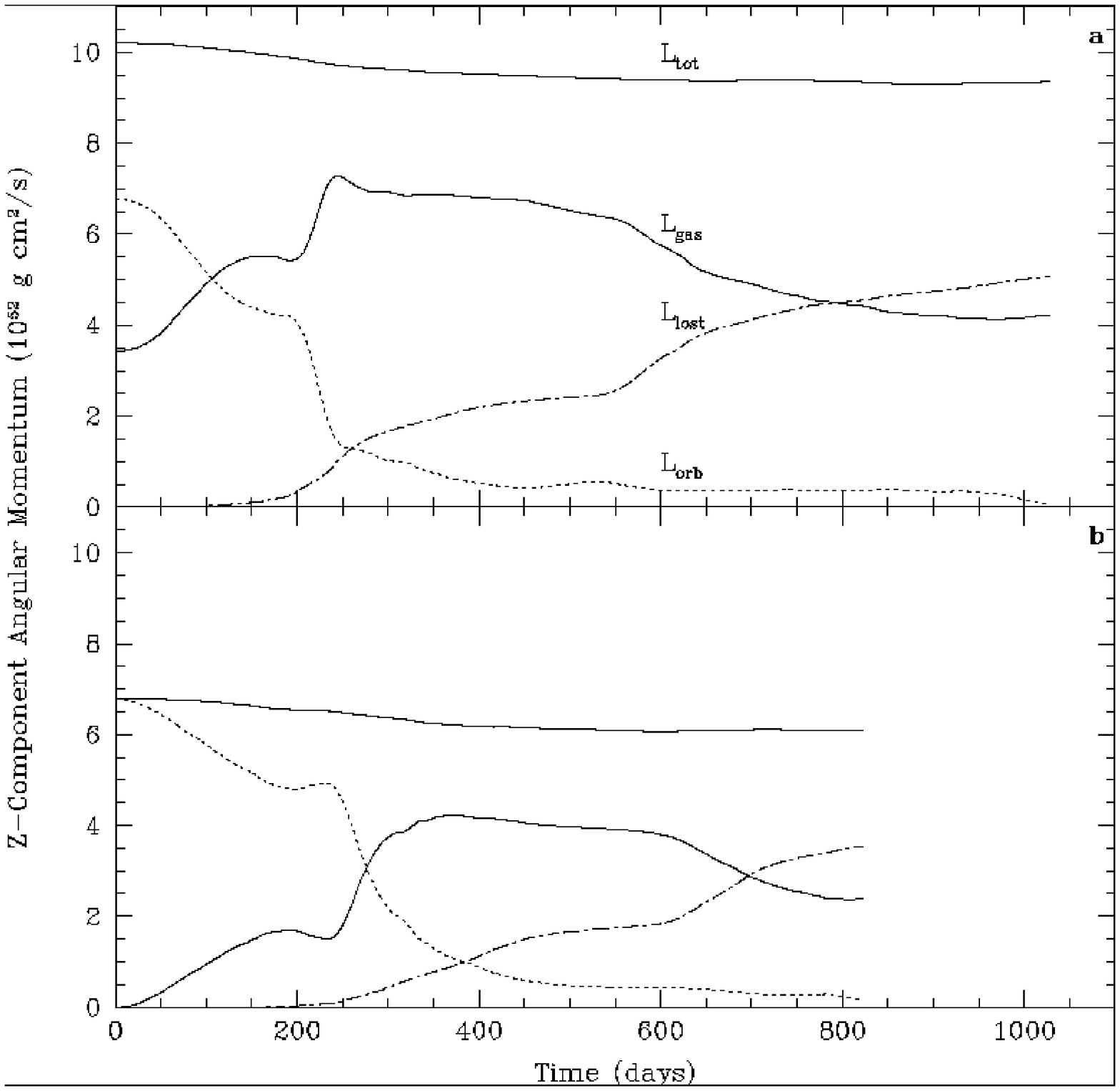}
\figcaption{The contributions to the z-component of the angular
momentum for a) simulation 1 and b) simulation 2 (no initial
rotation of the gas). The curves from top to bottom at 600 days are:
total, gas, lost from the grid, and orbital (carried
by the point masses) angular momentum. \label{ang1}}
\end{figure}

In the earliest phase of the evolution, gas from the near surface of
the giant star is gravitationally torqued up, removing energy and
angular momentum from the binary orbit in the process. (The energy and
angular momentum as a function of time are shown in Figures~\ref{erg1}
and \ref{ang1}.) We define this initial phase as the time interval
between the beginning of the simulation (a state approximating a
realistic initial binary orbit) and the time at which the separation
of the point masses is equal to the initial radius of the giant. In
our baseline simulation, this phase lasts 160 days.


During this initial phase, there is a period of fairly heavy mass
transfer from the giant to the companion. The matter stream is dragged
around behind the companion's direction of travel into an orbit until
the stream impacts itself, after which it quickly fills the
companion's potential well.

By the time the companion reaches the former surface of the giant, the
core and companion have transferred about one-third of their angular
momentum to the gas. Very little mass has been unbound at this point
though, and none has been lost. In addition, the energies in the
system have changed very little, as the gas surrounding the giant core
(which contains most of the energy) has not been significantly
affected by the distant influence of the companion.

\subsection{Rapid Infall Phase}

\begin{figure}[t]
\hspace*{1.3in}
\epsfxsize=9 true cm
\epsfysize=9 true cm
\epsffile{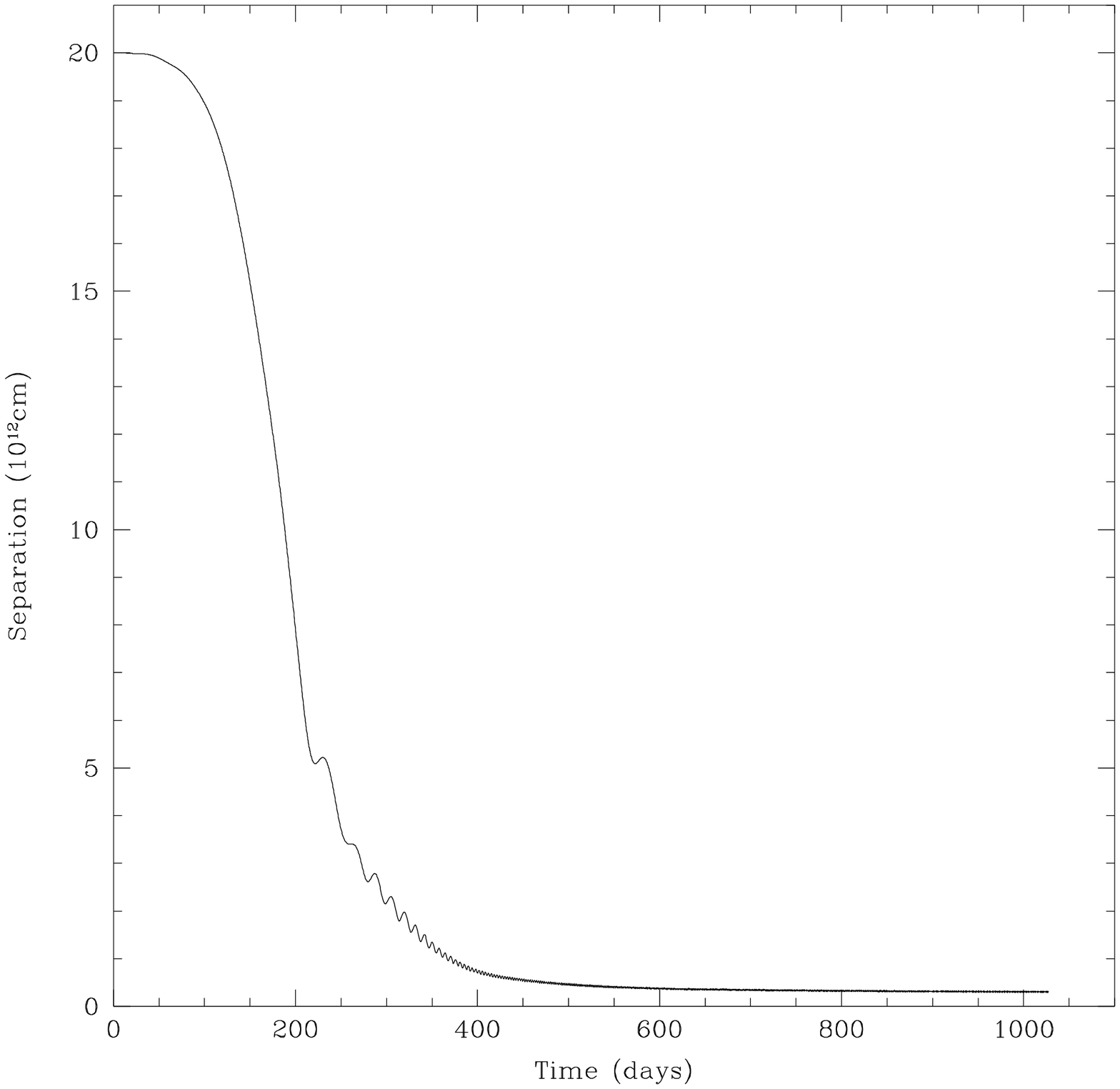}
\figcaption{The orbital separation between the two point masses in
simulation 1. \label{aa}}
\end{figure}

\begin{figure}[t]
\hspace*{1.3in}
\epsfxsize=9 true cm
\epsfysize=9 true cm
\epsffile{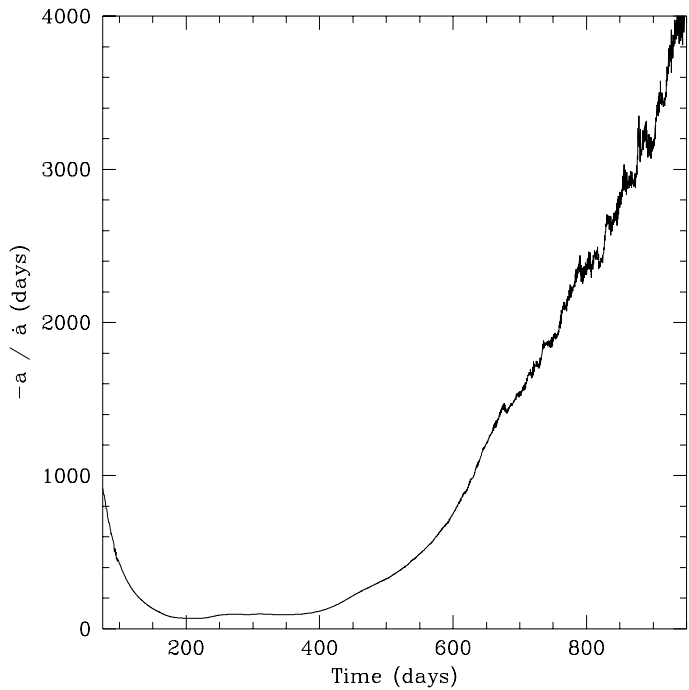}
\figcaption{The orbital decay timescale ($-a/\dot{a}$) for the point masses in
simulation 1. A time-averaged orbital separation
has been used in computing the timescale. \label{adot}}
\end{figure}

Once the companion has entered the envelope, which is still mostly
intact, the orbital decay rate of the binary ($-\dot{a}$) has almost
reached its maximum value (over a percent of the giant's radius per
day; see Figures~\ref{aa} and \ref{adot}). Over the course of the next
100 days the majority of the change in the core-companion separation
occurs. We define the end of this phase as the time at which the
orbital decay timescale begins increasing again, after holding at a
roughly constant value.  This occurs after approximately 400 days. In
this simulation, about 90\% of the remaining orbital angular momentum
is transferred to the gas, mostly during the early part of the phase
when the binary separation is decreasing most rapidly.


\begin{figure}
\hspace*{1.3in}
\epsfysize=18 true cm
\epsffile{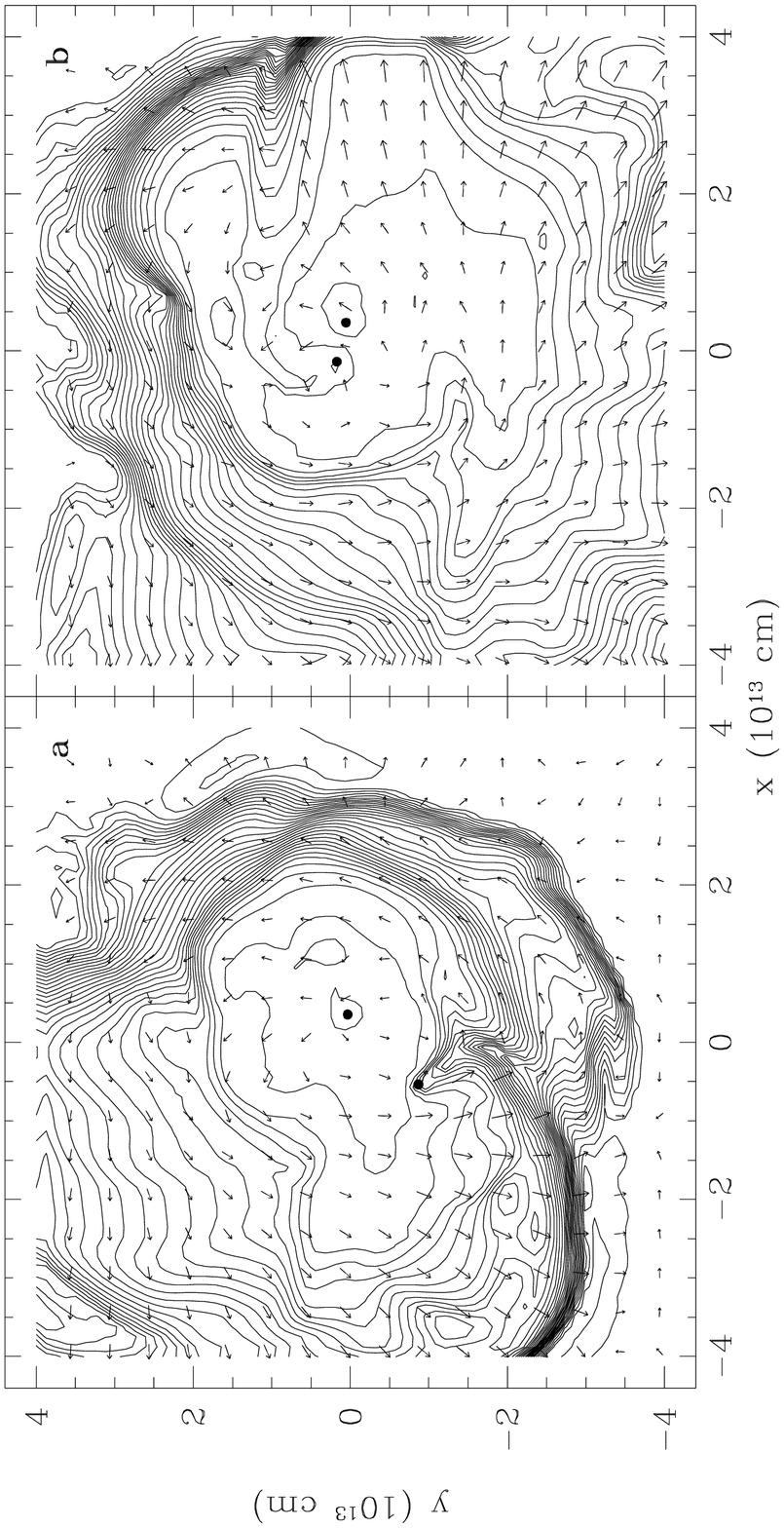}
\figcaption{Specific entropy contour plots at a) 170 days and b) 234
days. Entropy contours are at intervals of 0.01 in the logarithm.
Solid dots indicate the positions of the point masses. \label{entfig}}
\end{figure}

\begin{figure}[p]
\hspace*{0.5in}
\epsffile{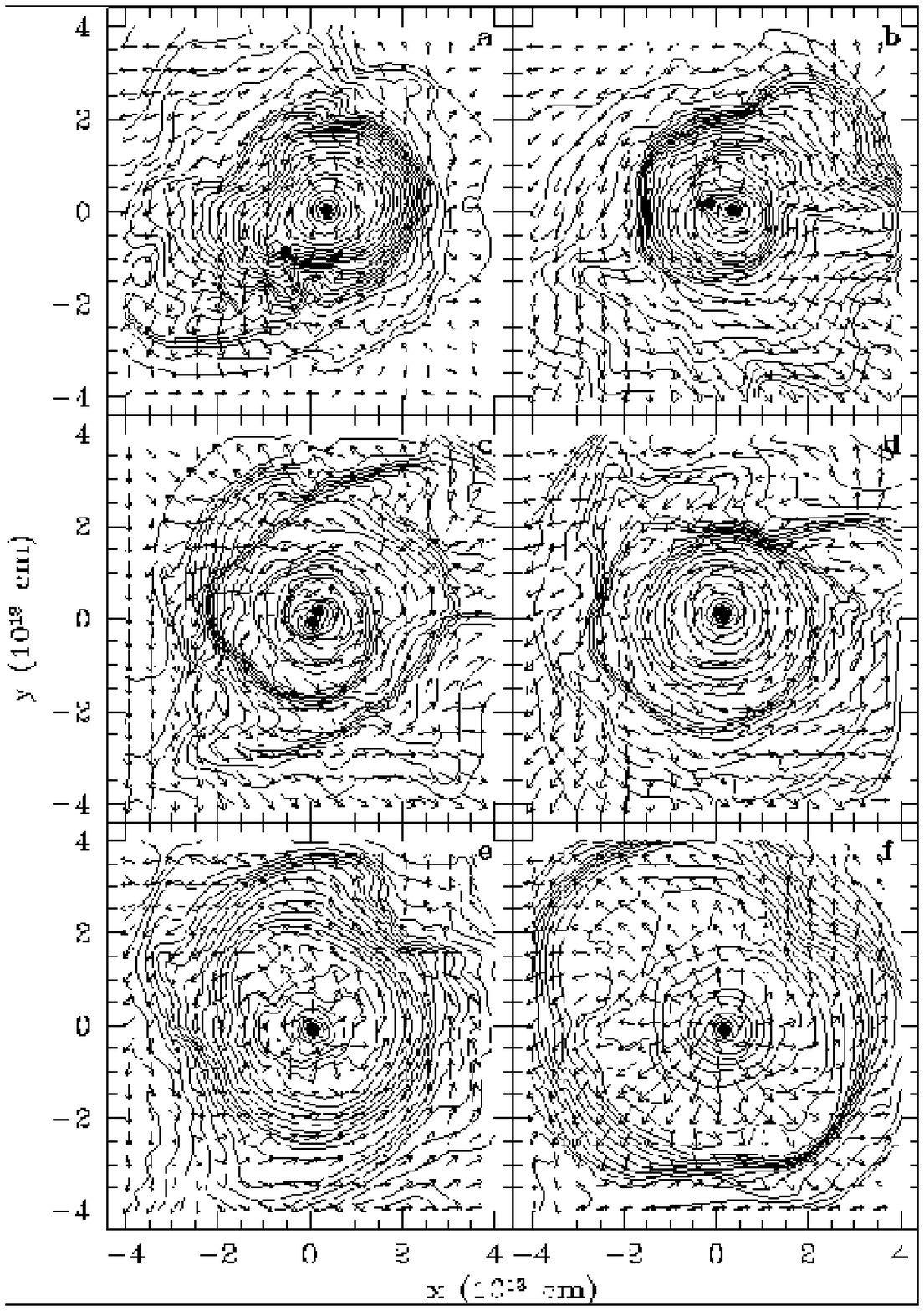}
\figcaption{A sequence of density contour plots in the orbital plane
during simulation 1. Contours are five per decade in density. Solid
dots indicate the positions of the two point masses. The velocity
vector field is scaled to the maximum value in each frame. From left to
right, and top to bottom, the age of the simulation at the time of
snapshots is a) 170 days, b) 234 days, c) 289 days, d) 372 days, e)
462 days, and f) 530 days. \label{rho6}}
\end{figure}

In the early portions of this stage, when the core-companion
separation is decreasing most quickly (on a timescale of about 100
days), the gas flowing past the companion is accelerated to supersonic
velocity. A bow shock is generated in this gas flow, directed more or
less in the direction of travel of the companion, as is visible in the
entropy contour plots in panel a of Figure~\ref{entfig}.  Later on
when the point masses have spiralled closer, weak spiral shock waves
can be seen in the density contours (see Figure~\ref{rho6}). At first,
only one shock is generated --- in front of the more rapidly moving
companion.  As a result, mass is pushed away from the giant in a very
asymmetric pattern. Even later, shock waves are generated in front of
both point masses. Near the point masses (before the waves have gone through
one-quarter of a turn), the shock waves are strongest,
in the sense that more entropy is generated in the gas.  When the
shock waves finally propagate into regions of lower density, the Mach
number of the gas increases, and they become more visible in the
entropy plots (as in panel b of Figure~\ref{entfig}). These shock
waves are the primary means for transporting angular momentum away
from the point masses during this phase.


\begin{figure}[p]
\hspace*{0.5in}
\epsffile{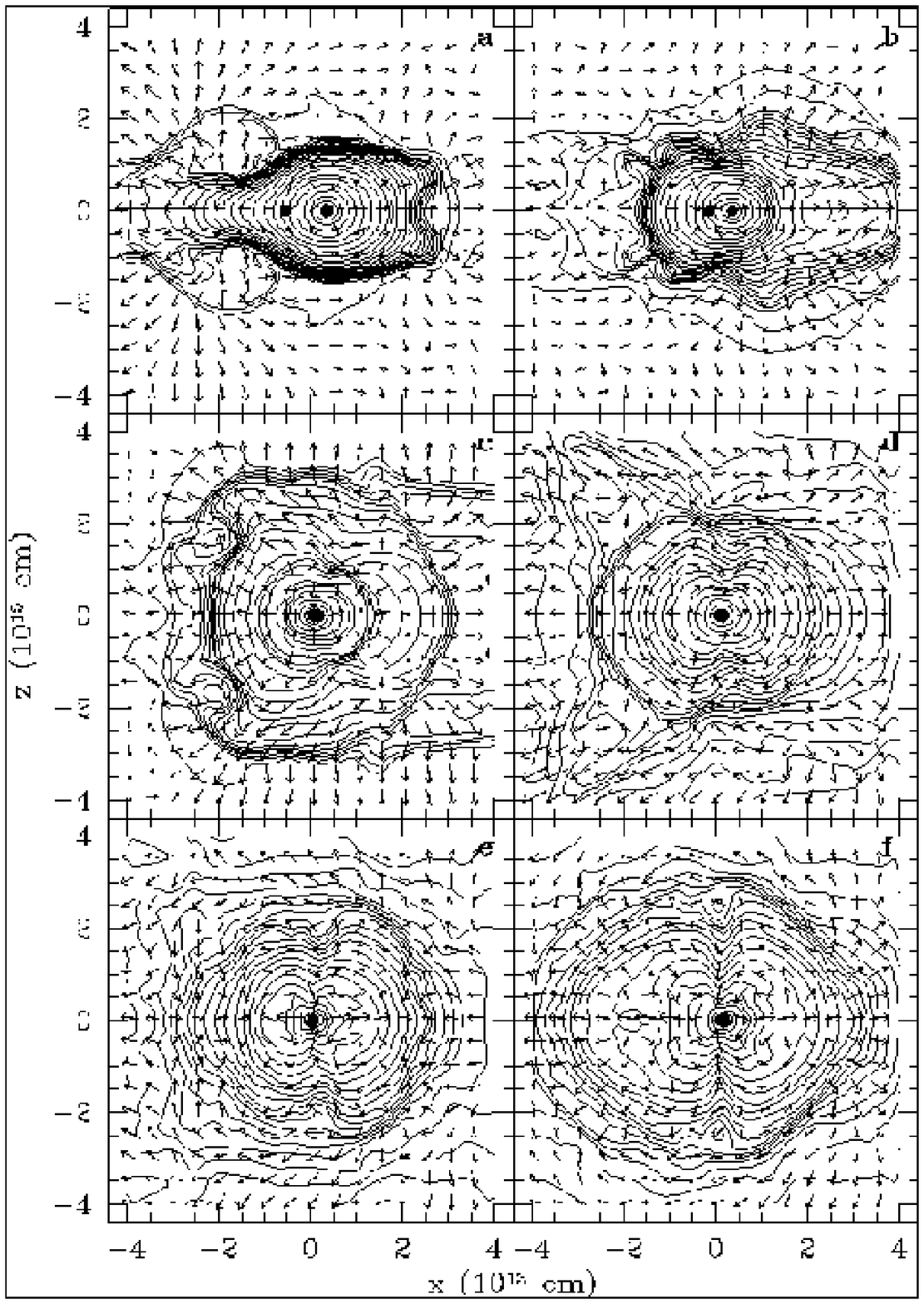}
\figcaption{A sequence of density contour plots perpendicular to the
orbital plane during simulation 1. The legend is the same as in
Figure~\ref{rho6}. \label{rho6xz}}
\end{figure}

Spiral shock waves continue to move outward from the point masses
throughout this phase. Because the period of the binary orbit is
considerably smaller than the orbital timescale of the envelope gas,
the spiral waves become tightly wound. The angular momentum input from
the shocks causes the mass distribution to elongate in the xz-plane,
as seen in panel c of Figure~\ref{rho6xz}.  The peak energy transfer
occurs shortly after this point as the companion begins to dissipate
energy in the high density gas at the core of the giant. At the same
time, the angular momentum transfer from the point mass orbits to the
gas is shutting off.  The gas near the two cores is torqued up, but
only a very small region ($\approx 10^{12}$ cm in diameter)
following the companion's spiral wave reaches angular velocities near
corotation.  At the very end of the phase, the spiral shocks are very
weak everywhere except in the vicinity of the point masses.  Outside
this region, the gas motions are close to radial, which is the
direction normal to the spiral waves. A thick disk-like structure
begins to take shape towards the end, and a low density region forms
along the rotation axis (see panel d of Figure~\ref{rho6xz}).

\begin{figure}
\hspace*{-0.5in}
\epsffile{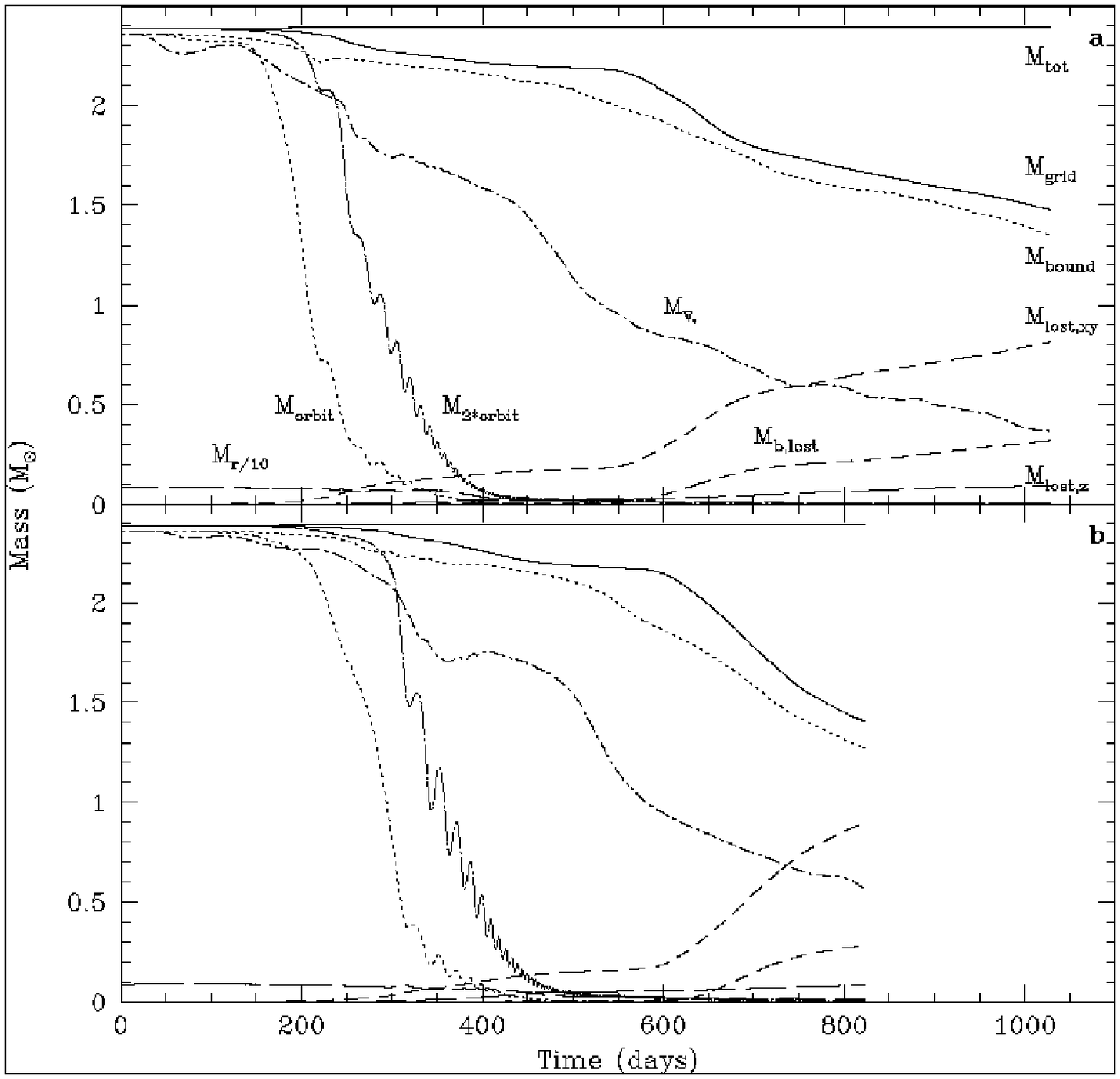}
\figcaption{Tracers of gas mass for a) simulation 1 and b) simulation
2 (no initial rotation of the gas). The curve labels are as follows:
$M_{tot}$, total mass; $M_{grid}$, mass remaining on the grid;
$M_{bound}$, mass on the grid that remains bound; $M_{V_{*}}$, mass
in the original volume of the giant; $M_{r/10}$, mass within a
distance of one-tenth the original stellar radius from the core of the
giant; $M_{orbit}$, mass within a circular orbit the size of the
current separation of the point masses; $M_{3*orbit}$, same as
$M_{orbit}$, but with three times the orbital separation;
$M_{lost,xy}, M_{lost,z}$, mass lost from the grid in the radial
direction and in the z-direction; and $M_{b,lost}$, bound mass lost
from the grid. \label{mass1}}
\end{figure}

The peak energy transfer from the orbits of the companion and the core
to the gas occurs when the two particles are deep in the potential
well of the system where more energy is available per unit change in
separation. Very little of this energy has been converted into kinetic
energy of the gas though. From Figure~\ref{mass1}, we infer that
most of the energy lost from the orbit is used in powering a
relatively slow expansion of the mass closest to the core and
companion. At the end of the phase there is little mass ($< 0.03$
\msun) within a distance of several times the orbital separation from
the point masses --- more than half of the original content has been
pushed outwards. On the other hand, the amount of mass within the original
radius of the star has decreased by about 30\%. This trend
continues into the next phase of the evolution.

\begin{figure}[t]
\hspace*{1.3in}
\epsfxsize=9 true cm
\epsfysize=9 true cm
\epsffile{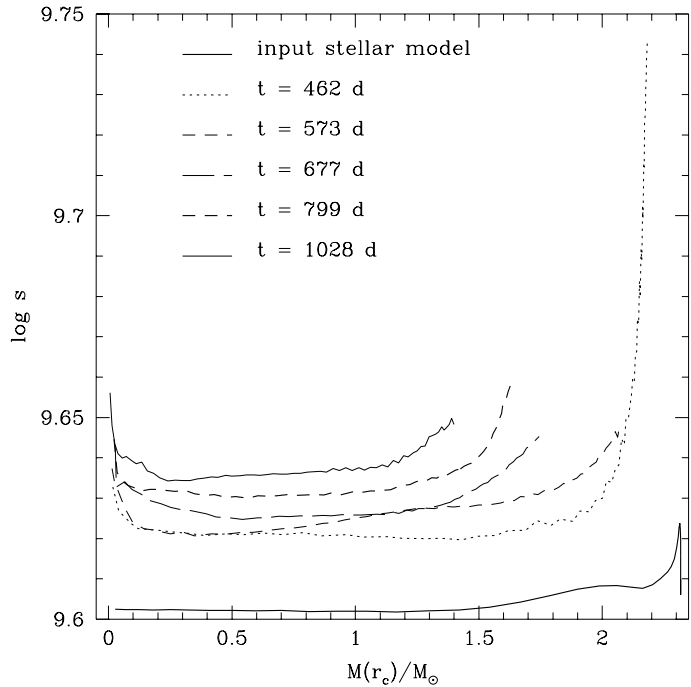}
\figcaption{Specific entropy profiles as a function of mass within cylindrical
shells at various times during simulation 1. The cylindrical radius
$r_{c}$ is measured from the center of mass of the two point masses.
\label{sprof}}
\end{figure}

\begin{figure}[t]
\hspace*{1.3in}
\epsfxsize=9 true cm
\epsfysize=9 true cm
\epsffile{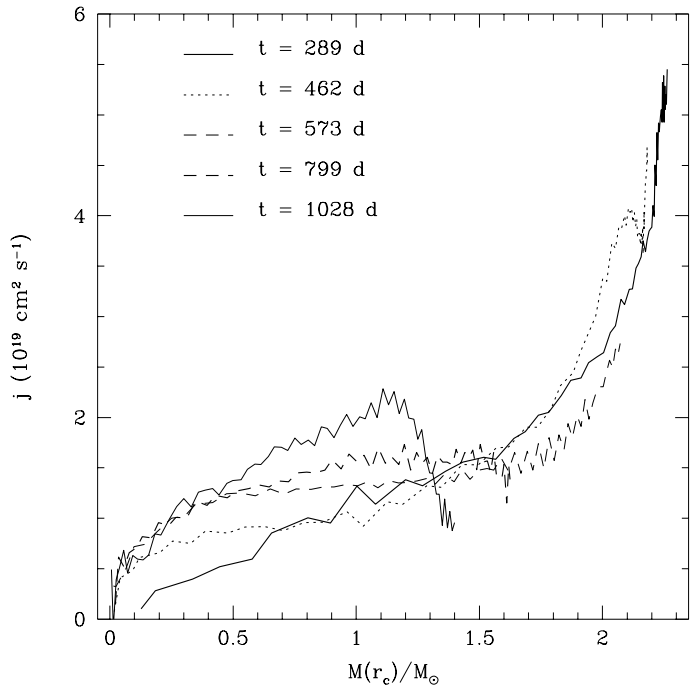}
\figcaption{Specific angular momentum profiles as a function of mass
within cylindrical shells at various times during simulation 1. The
cylindrical radius $r_{c}$ is measured from the center of mass of the
two point masses. The profile for an age of 677 days was not plotted
due to its similarity to that for 799 days. \label{jprof}}
\end{figure}

Through the midpoint of this phase, the specific entropy profile
increases monotonically with cylindrical radius (see
Figure~\ref{sprof}). By the end of the phase, a small amount of mass
($\approx 0.1$ \msun) near the core has developed a profile that
decreases outward, which indicates convective instability. This region
is maintained during the remainder of the simulation, as gas motions
apparently do not have time to redistribute the entropy. (In the next
phase of the evolution, we do see some evidence of slightly higher
entropy blobs in the vicinity of the point masses. This seems to
indicate the presence of convection, although the motions of these
blobs tend to keep them fairly close to the orbital plane.) At this
point, we also find the first indications of circulation perpendicular
to the orbital plane with timescales on the order of a year.  This
entropy profile around the point masses comes into being only when they
take up close orbits around each other, and are able to
repeatedly shock the gas. At about the same time, the specific angular
momentum of gas in the vicinity of the point masses also drops by a
large amount (see Figure~\ref{jprof}), while much of the gas outside
this region gains significantly. This appears to be the result of the
flattening of the giant.


\begin{figure}
\hspace*{1.3in}
\epsfxsize=9 true cm
\epsfysize=9 true cm
\epsffile{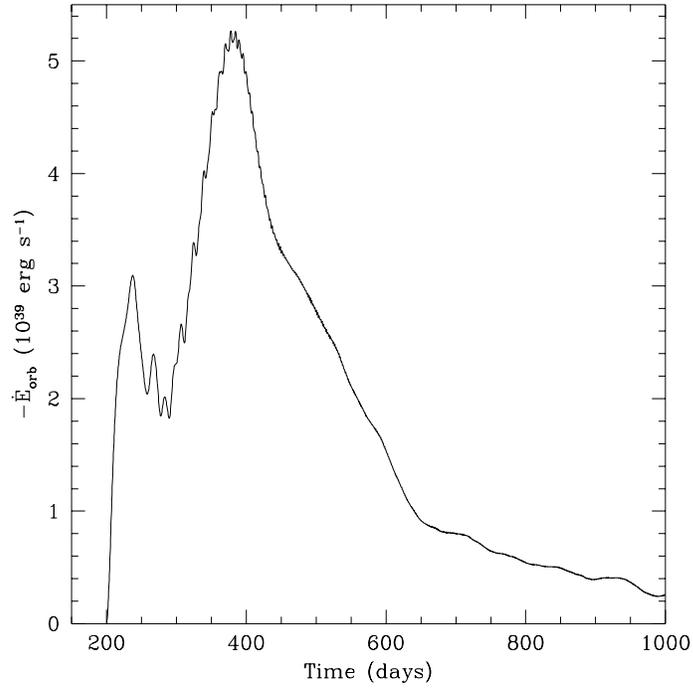}
\figcaption{The energy dissipation rate from the point masses orbits in
simulation 1. \label{edot}}
\end{figure}

At the end of this phase, the point masses are in close, relatively
stable orbits about each other. The center of mass of the orbiting
point masses has received most of its ``kick'' from the ejected
material also --- a speed amounting to just 3 km s$^{-1}$. The cores
begin oscillating around the system's center of mass due to an
asymmetry in the gas distribution, probably resulting from the spiral
outflow pattern.  The majority of the gas is still found within the
original volume of the giant, but it continues to flow outwards as a
result of the energy input from the core and companion.
The energy transfer rate (shown in Figure~\ref{edot}), which reached a
peak value of $1.3 \times 10^{6} \lsun$, has declined by about half.
Further expansion of the gas is facilitated through the use of a
larger proportion of energy taken from the internal energy of the gas.
The removal of mass from the vicinity of the point masses was
initially driven by spiral shocks, but at the end of this phase the
reduced mass near the cores has considerably reduced the efficiency of
angular momentum transport by them.  Regardless, the gas is
distributed primarily in the orbital plane of the binary as a result
of the earlier action of these waves.

\subsection{Envelope Ejection Phase}

\begin{figure}
\hspace*{1.3in}
\epsfxsize=9 true cm
\epsfysize=9 true cm
\epsffile{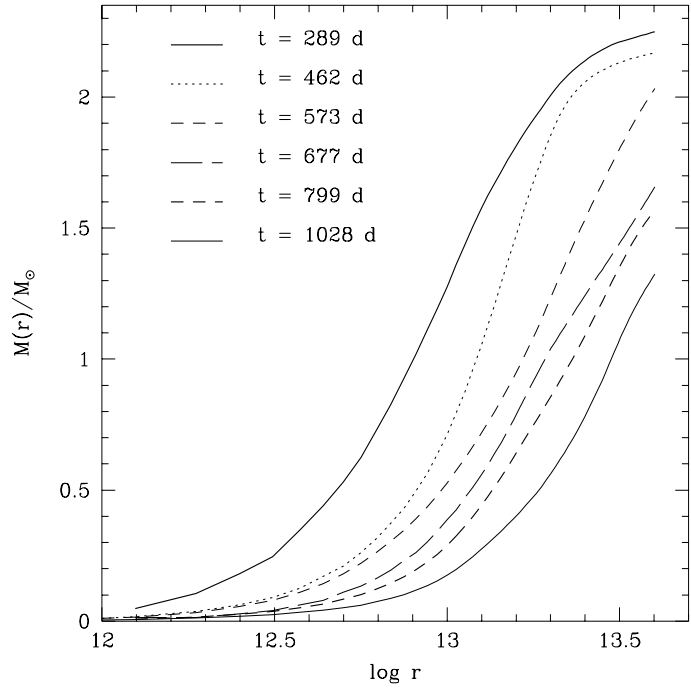}
\figcaption{Mass-radius profiles for various times during simulation
1. The radius is measured from the position of red giant core.
\label{mrd}}
\end{figure}

At the beginning of this phase, the cores have completed the majority
of the orbital decay with the separation reaching $\sim 5 \times
10^{11}$ cm.  The orbital decay timescale increases dramatically
beyond this point.  This is primarily because the mass in the
immediate vicinity of the two cores has declined to $0.025 \msun$.
There was originally about $0.1 M_{\odot}$ of gas in this region of
the red giant initially --- amounts this small are typical of the
mass-radius profiles of red giants (see, for example, Figure~6 of
Yorke at al. 1995, as well as Figure~\ref{mrd}).  We note, however,
most of the mass loss from the original volume of the giant (which we
will call the ``ejection'') does not occur until after the cores have
established themselves in close orbits about each other.

The gas heated by close interaction with the two cores begins
expanding away, and continues expanding through the remainder of the
simulation, as shown in panels e and f of Figure~\ref{rho6}. The
rotational pattern associated with the disk-like structure seen in
panel d is quickly overwhelmed by the outflow. The expansion of the
gas is driven by continued energy input from the point mass orbits,
and is responsible for the movement of most of the remainder of the
giant's mass out of the star's original volume. This energy input also
appears to be the cause of the circulations that develop perpendicular
to the plane of the orbit within about $10^{13}$ cm of the point
masses (see especially panels e and f of Figure~\ref{rho6xz}),
confirming earlier work by Taam \& Bodenheimer (1991) in two
dimensions and Terman \& Taam (1996) and Rasio \& Livio (1996) in
three dimensions. The majority of the gas on the grid is still to be
found within about one giant radius of the orbital plane.

By 530 days (panel f in the density plots), two thin cones having
strong inflow are set up along the rotation axis.  These inflows
complete the circulation pattern, partially replenishing the gas
surrounding the point masses that is being driven out. The strongest
outflow is to be found close to the orbital plane, where gas can be
accelerated to faster than the local sound speed.  From
Figure~\ref{mass1} it is apparent that there is a net outflow, as can
also be gathered by the fact that the strongest inflow involves
relatively low density gas along the rotation axis. At later times,
the outflow becomes more confined to the orbital plane ($\pm 5 \times
10^{12}$ cm).

\begin{figure}
\hspace*{1.3in}
\epsfysize=18 true cm
\epsffile{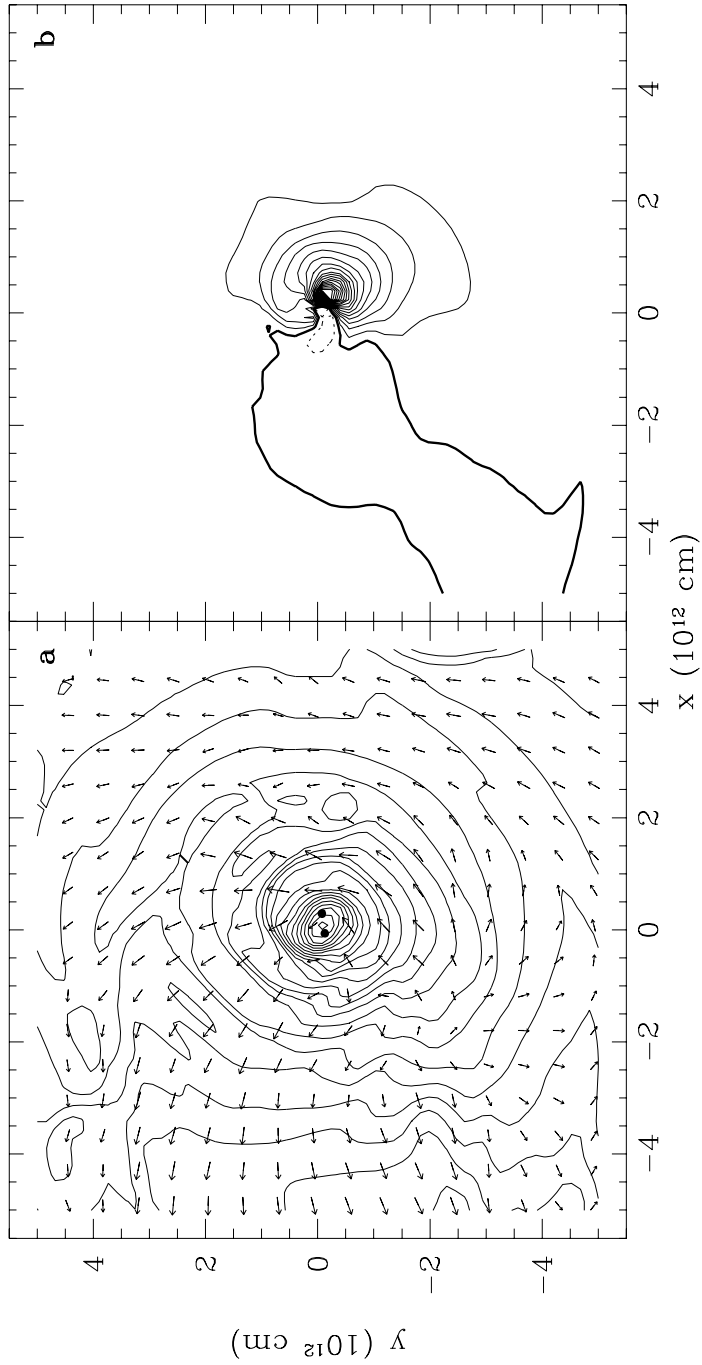}
\figcaption{Density (panel a) and angular velocity (panel b) contours
in the orbital plane for the innermost subgrid at 677 days. Density
contours are 5 per decade. Angular velocity contours are at intervals
of 2.5\% of corotation. The dark solid line indicates the zero
rotation contour.  The positions of the point masses are indicated in
on the density contour plot. \label{rhoomeg}}
\end{figure}

At the end of the simulation, over 90\% of the gas was pushed out of
its original volume, and mass was still being removed from the volume
at a rate of approximately 0.3 \mpy.  About 45\% of the gas had been
pushed off the edge of the grid, although almost all of what remains
is still bound (although marginally) to the system.  At the end of the
simulation, a significant amount of bound mass had been lost off the
grid as a result of the expansion of the gas, making continued
calculations unreliable.  As was found at the end of the rapid infall
phase, the majority of the gas does not have angular velocity more
than 10\% of corotation. As seen in Figure~\ref{rhoomeg}, a small
amount of mass is maintained close to corotation very near the
orbiting point masses, but the angular momentum transfer is
negligible.

In the end, the point masses have a separation of $4.42 \rsun$. At
late times, the fractional decrease in the orbital separation of the
cores should equal the fractional decrease in their gravitational
potential energy since there is little envelope mass within the binary
orbit. We do find that the energy loss and orbital decay timescales go
in lockstep. By the end of the simulation, the orbital decay timescale
had increased to about 12 years. The mass loss rate for the system
(computed from the amount of bound gas) has remained roughly constant
over most of the simulation, meaning the mass loss timescale has
decreased to about 6 years by the end of the simulation.  A comparison
of the timescales indicates that the entire common envelope should be
ejected. As can be seen from Figure~\ref{mass1}, the gas near the two
cores is cleared out within about 100 days after the end of the rapid
infall phase, and at the end of the simulation only about $0.37 \msun$
remains in the original volume of the giant.

\bigskip

\subsection{Comparison Simulations}

In all but the last of the following sequences, the companion was started on a
circular orbit at a distance of $2.0 \times 10^{13}$ cm away from the
core of the giant. In all of the sequences but 2 and 5, the
envelope of the giant star was in synchronous rotation with the
companion's orbit. In Table~2, we summarize various characteristics of
the system at the end of the simulation: final separation of the point
masses, their orbital period, the recoil velocity for the center of
mass of the point masses, and the efficiency of envelope ejection
$\alpha_{CE}$. [We choose to use the definition 
\[ \alpha_{CE} = \frac{\Delta E_{bind}}{\Delta E_{orb}} \]
(Tutukov \& Yungelson 1979), where $\Delta E_{orb}$ is the change in
orbital energy of the point masses (which is the difference in the sum
of kinetic plus potential energies --- in the final state this was
computed for the two point masses, but for the initial state, this was
computed using the masses of the companion point mass and the giant),
and $\Delta E_{bind}$ is the binding energy of the mass ejected, as
determined from the initial giant model. The mass ejected was taken to be
the mass lost from the main grid plus unbound mass remaining on the
grid minus bound mass lost from the main grid.] Many details of the
sequences are similar, so we will concentrate on the differences in
the final orbital parameters and the mass-loss evolution.

\subsubsection{Sequence 2: 3 \msun Giant, 0.4 \msun Companion,
Nonsynchronous Rotation}

In this simulation, the initial conditions were chosen to be identical
to those of the baseline simulation, except that the giant envelope
was not rotating. The expectation was that the larger relative velocity
between the companion and the envelope of the giant would result in a
greater deposition of the companion's orbital energy and angular
momentum in the envelope.

A comparison of the mass tracers for the sequences
(Figure~\ref{mass1}) indicates that the timescale for the companion to
enter the envelope of the giant is actually slightly longer for the
nonsynchronous case.  This is a result of less efficient angular
momentum transfer for the nonsynchronous case --- in order to begin to
enter the envelope, the companion's orbit must develop eccentricity,
as can be seen in Rasio \& Livio's (1996) calculation, and in ours. In
this simulation, the companion's gravitational torque acts for smaller
amounts of time on any particular element of the surface of the
giant's envelope.

Mass is removed from the star's volume and from the grid at
approximately the same rate for both simulation 1 and 2. We do not
find any observable characteristics of the simulation that would allow
us to determine whether the giant in such a binary was out of
synchronism at the beginning of the interaction.

\subsubsection{Sequence 3: 5 \msun Giant, 0.4 \msun Companion}

A simulation was run with a $5 \msun$ AGB star to gauge the effects of
different giant star mass (with a more massive core and a more massive
envelope) on the evolution.  Since the $5 \msun$ giant had nearly the
same initial radius as the $3 \msun$ giant and the initial orbital
separation was the same as in sequence 1, the giant overfilled its
Roche lobe by a smaller margin.  This sequence was not followed as
long as the baseline simulation (891 days, as opposed to 1027 days for
sequence 1), but at the end of the simulation the core-companion
separation was found to be converging to approximately the same value.

\begin{figure}
\hspace*{-0.5in}
\epsffile{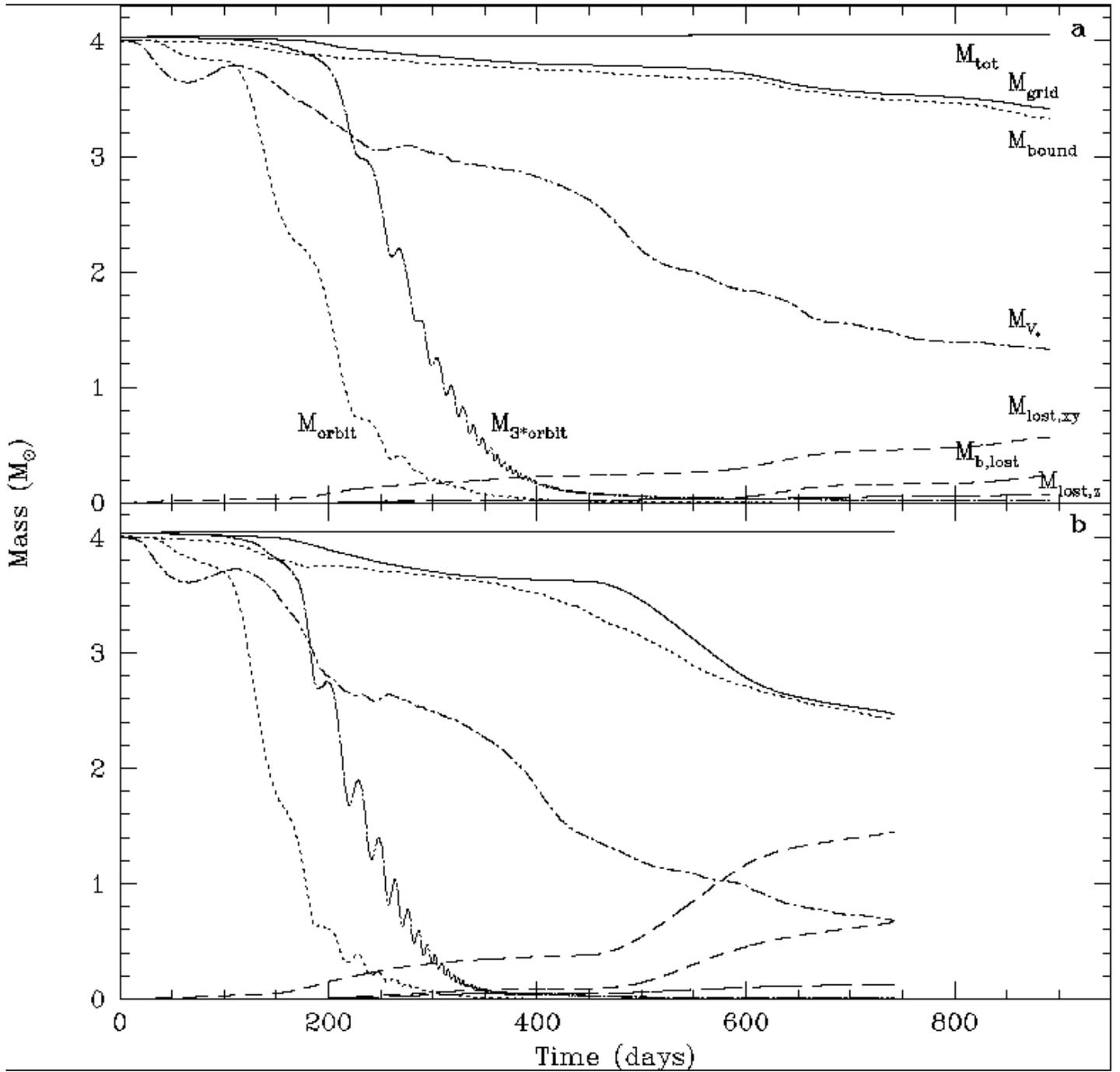}
\figcaption{Tracers of gas mass for a) simulation 3 and b) simulation
4. The curve labels are the same as found in Figure~\ref{mass1}.
\label{mass2}}
\end{figure}

Although the binding energy of the envelope is larger, gas is lost
from the original volume of the star and from the grid at nearly the
same rate as in the baseline simulation, as shown in
Figure~\ref{mass2}. The implications are that the energy released from
the orbit in these two simulations goes towards almost completely
clearing out the region around the orbiting point masses, with the
remainder (probably the majority) going toward expansion of the
envelope in the orbital plane. If, as appears to be the case, the
orbital decay of the point masses is physically slowing (as opposed to
slowing due to numerical effects), we would expect that the envelope
of the giant would not be completely ejected. At the end of this
simulation the orbital decay timescale is about 4 years, which is
roughly half what it was at the comparable point in simulation 1. In
addition, the orbital decay timescale is increasing at roughly half
the rate it was in simulation 1. The mass-loss timescale at the end of
the simulation is much longer ($\approx 27$ years) than the orbital
decay timescale, further suggesting that the envelope of the giant
might remain close enough to the point masses to affect their orbits
further.

The inability of the companion to eject the more massive envelope is
also reflected in the efficiency factor $\alpha_{CE}$ --- for this
simulation we find that is only 24\%, compared to $\approx$ 40\% for
sequences 1, 2, and 4. However, as discussed in \S~\ref{numerical}, a
different algorithm for the smoothing length in the gravitational
potential leads to considerably more mass ejection and a higher
efficiency factor, so these results for this simulation should be
treated with care.

\subsubsection{Sequence 4: 5 \msun Giant, 0.6 \msun Companion}

In this simulation, we again find that the companion's orbital
decay has stopped at approximately the same distance from the giant
core as it did in the baseline simulation. The flat portion of the
initial mass-radius profile of the $5 \msun$ giant had very nearly the same
spatial extent as that of the $3 \msun$ model, which may explain the
similarity if indeed the orbital decay decelerates when the two point
masses begin orbiting within this region.

At the end of the simulation (which lasts 742 days), the orbital decay
timescale has almost reached 6 years, indicating that the mass near
the point masses has been efficiently removed. The timescale is
actually increasing faster than in simulation 1.  In examining the
mass tracers, we find that the same {\it fraction} of the giant
envelope is removed in half the time of the baseline simulation.  At
the end, the mass-loss timescale as judged from the bound mass is
about 17 years and decreasing.  However, most of this gas 
is marginally bound, and it is sufficiently far away that
it does not influence the point-mass orbits in the short term. The
mass in the original giant volume is decreasing considerably more
quickly than in simulation 3.

\subsubsection{Sequence 5: evolved 5 \msun Giant, 0.6 \msun Companion,
Nonsynchronous Rotation}

As a test of the effects of the evolutionary state of the giant on the
evolution, we conducted this simulation in which the 5 \msun giant was
in a more evolved state. The input giant model had a radius of $2.46
\times 10^{13}$ cm, as opposed to $1.32 \times 10^{13}$ cm for the
previous runs with 5 \msun giants. The carbon-oxygen (CO) core was
0.87 \msun, and 0.57 \msun for the previous runs. (The point mass had
a slightly lower mass than for the previous runs because the point
mass includes all the mass of the star up to the size scale of about
the smallest zone.  Even though the CO core is larger, the overall
expansion of the giant, reduces the mass relegated to the point
particle.) In addition, the orbit of the companion was moved out to a
distance of $3.7 \times 10^{13}$ cm.

\begin{figure}[t]
\hspace*{1in}
\epsfxsize=12 true cm 
\epsffile{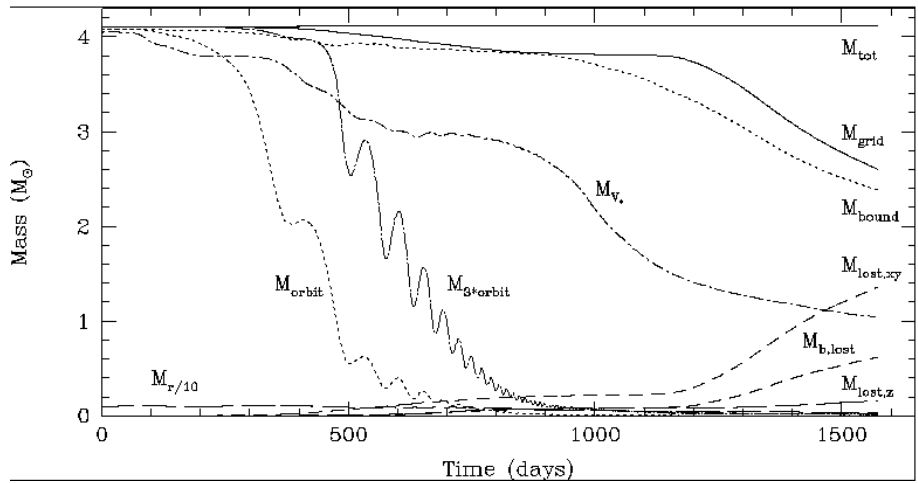}
\figcaption{Tracers of gas mass for simulation 5. The curve labels
have the same meaning as in Figure~\ref{mass1}. \label{mass3}}
\end{figure}

The simulation was run for 1574 days, and the final separation of the
point masses in this simulation was approximately twice that of the
previous simulations. The mass tracers are shown in
Figure~\ref{mass3}. Because this could have been related to our
numerical algorithm, we explored this possibility, as described in the
next section. The result was that the smoothing length was found to
affect the final separation, but not the efficiency of envelope
ejection, which was found to remain roughly constant near 53\%.

The spiral-in phase lasts for the first 300 days in this simulation,
while the rapid infall phase also lasts roughly twice as long as in the
other simulations (ending at about 800 days). At the end, the mass
loss timescale has increased to 16 years, while the orbital decay timescale
is at 7 years, and both are increasing. The final outcome of the
simulation wil probably be ejection of the envelope since the gas
within the giant's volume is being pushed out quickly.

\subsection{Numerical Effects}\label{numerical}

The later portions of the envelope ejection phase were not followed by
Rasio \& Livio (1996) because the local spatial resolution for their
SPH gas particles near the cores was comparable to the size of the
binary orbit. In our simulations, limitations on spatial resolution
are present in the zone spacing on the grid the point masses occupy,
and in the smoothing length applied to gravitational interactions
between point masses and gas.

With regards to the effects of the smoothing length in the gas-to-point-mass
potentials, the main questions are how critical the formulation of the
potential is, and how physically realistic it is.  The potential we
have used crudely mimics the effects of constant density mass distributions
in each gas zone, since the force due to a given zone is reduced when
the point mass falls within the zone.  The point masses are not
strictly points, although their physical sizes (in the case of compact
objects, including the cores of giant stars) are more than an order of
magnitude smaller than the smoothing length.  However, relatively
small amounts of matter accreted onto these stars will cause the cores
to partially fill their respective Roche or tidal lobes, creating
configurations that are much larger (see Hjellming \& Taam 1991) and
comparable to the zone size. We expect that the smoothing length only
becomes important towards the end of the simulations when the point
mass orbits are small, meaning that the point masses orbit within and
primarily interact with only a handful of zones.

To test the effects of the smoothing length, we allowed it to decrease
once the point mass separation reached a distance of three times the
smoothing length of one point mass. To an extent this is physically
realistic if the ``true'' size of the point masses are on the order of
the tidal radius, since the tidal radii will decrease as the orbital
separation decreases. In making this change for the configuration of
sequence 5, we find that the final orbital separation decreases from
about 8.9 \rsun to 5.3 \rsun. We also ran the same simulations with a
reduction in the number of subgrids from two to one to check the
relative importance of the grid resolution itself. For a fixed
potential smoothing length 1.5 times the smallest zone size (twice the
size of sequence 4), we found the final separation was
about 16.9 \rsun. When the smoothing length was fixed equal to the
smallest zone size, the separation became about 14.9 \rsun. With the
adjustable smoothing length algorithm, the final separation reached
8.1 \rsun.

For simulations with additional subgrids, we expect the computed final
separations would converge on a value.  However at our present level
of resolution, we are unable to specify exactly how close our two
point masses approach each other in the end. We should consider
whether we can put realistic limits on the final separation. It is
safe to say that the separations derived from our fixed smoothing
length simulations are good upper limits since higher resolution would
tend to increase the amount of energy dissipation and angular momentum
transfer. We applied the newer smoothing length algorithm to
simulations 1, 3, and 4, and found that the final separation did
decrease in all cases, but by no more than 25\%. The additional energy
input into the gas leads to to the unbinding of considerably more mass
and more mass loss off of the grid, while the amount of bound mass
loss off the grid remains about the same. For simulation 3, the
additional decrease in orbital separation and the resultant mass loss
might be enough to eject the envelope.

The fact that the orbital separations still end up being nearly the
same with the new smoothing length algorithm may indicate that the grid
resolution has become the most important factor. We must conclude that
it is unclear exactly how close the adjustable smoothing length
algorithm brings us to the real physical separation.

There are a few final points we wish to make. First, the first two
stages of the common envelope evolution are unaffected by this
question, and most of the details of the envelope ejection phase will
not change. For example, the efficiency of envelope ejection probably
will not change much since the final separation is implicitly taken
into account in the calculation. In fact, for several test
versions we ran of sequence 5, the efficiency only varied by a few
percent, despite differences in the final orbital separation and
orbital energy of more than 50\%. Second, in our simulations with
fixed smoothing length, there is considerable eccentricity of the point mass
orbits. This shows up in the rapid variations of the point-mass kinetic and
core-companion gravitational energies in the late stages of the
simulations. When we ran the same simulations with the smoothing
length allowed to shrink with the orbit, the eccentricity is damped out.
We believe this damping will be observed in higher resolution computations.

The anonymous referee points out that our conclusions about the degree
of corotation of the gas immediately surrounding the the two point
masses may be affected by resolution effects. Rasio \& Livio (1996)
found that reducing the resolution (by reducing the number of
particles) suppressed corotation of the gas at the end of the
simulation. An examination of the present work and that of Rasio \&
Livio indicates that the two are not inconsistent with regards to the
rotational state of gas farther from the cores. While the differences near
the cores may be a function of numerical resolution, the
results of the two studies indicate that the region of corotation is
rather small in spatial extent (of the order of the binary separation)
and negligible in mass, and hence dynamically unimportant.

\subsection{Outcomes of Common Envelope Evolution}

There are a number of physical effects that have not been included in
our models of the common envelope evolution, but which may affect
either our results, or our inferences about the eventual outcome of
the interaction. In the following, we briefly discuss several.

As discussed in the previous section, we have modeled the core of the
red giant and the companion star as point masses. As a result we
neglect aerodynamic drag effects caused by the finite size of the
stars and the gas they have accreted. This would only tend to
influence the results during phases in which the density of gas
impinging on the stars significantly affects the momentum of the
stars. So, this probably does not affect the orbital evolution of the
stars beyond the ends of our simulations, but could influence the rate
at which the companion initially falls through the giant's envelope,
or the orbital decay before too much of the envelope is ejected from
the vicinity of stars. Even so, for objects traveling at supersonic
speeds through a gas, aerodynamic drag is found to be small in most
cases compared to the gravitational drag (Kley, Shankar, \& Burkert
1995), which we do model. Thus, we do not expect aerodynamic drag to
play a large role.

It is also certain that radiation transfer becomes important at times
beyond the end of the simulations. As the gas density of the nebula
drops, the decrease in optical depth allows the gas to cool, removing
internal energy that could have been used to unbind it. In addition,
radiation from the white dwarf core of the red giant could help to
power a wind that could assist in removing the remnants of the
envelope. If gas is able to cool and contract back towards the cores,
it could result in additional reduction of the core-companion
separation. A rough calculation applying an analytic expression for
bound-free opacities to the density and temperature distribution
indicates that radiation diffusion timescales from the vicinity of the
point masses are on the order of a thousand years at the end of the
simulation. It seems clear that gas must rarefy considerably before
radiation losses become important.

\section{Discussion}

In this paper the hydrodynamical evolution of a binary system
consisting of an AGB star and its main sequence companion has been
followed from the initial to the late stages of the common envelope
phase using high resolution numerical techniques.  During the
evolution, spiral density waves clearly show the coupling between the
gas and the double core.  The shrinkage of the companion orbit is very
rapid, lasting only about 0.7 year.  The orbital decay decelerates
dramatically at an orbital separation of $\lapprox 7 \rsun$, although
this may be due in part to the effects of finite spatial resolution.
The double core is briefly embedded in a slowly expanding
differentially-rotating pressure-supported disk, which remains loosely
bound to the two cores.  The matter within the original volume of the
common envelope is spun up to only a fraction of corotation with the
point masses, but energy input from the point mass orbits leads to
mass removal at a rate of $\sim 0.3$ \mpy at the end of the
simulation.

The evolution to the late stages was followed for about an order of
magnitude longer in time than Rasio \& Livio's (1996) SPH calculation.
The results of our studies show that the late phases of the
hydrodynamical evolution are not affected by the initial spiral-in of
the companion into the giant star.  That is, it is found that the
degree to which the components are out of synchronous rotation at the
onset of the common envelope phase only affects the initial
development.  A slightly less rapid infall of the companion into the
red giant envelope results.  The final orbital separation of the
remnant binary immediately following the ejection of the common
envelope has not yet been reached in our simulations because of
inadequate spatial resolution. For our first four simulations, a
period of one day is found to be the upper limit. With a more evolved
giant, the final period appears to increase.

The results of the numerical simulations suggest the successful
ejection of the common envelope for a binary consisting of a $3\msun$
giant with a $0.7 \msun$ core and a $0.4 \msun$ companion.  On the
other hand, the system may or may not merge for a
higher mass giant of $5 \msun$ (with a $1 \msun$ core) with the same
companion.  The tendency toward merger for higher mass systems
reflects the larger mass that must be ejected and the greater
potential well from which the common envelope must escape.  For more
advanced stages in the evolution of the giant, a greater fraction of
the mass is situated at larger distances from the degenerate core so
that the energy requirements for ejection can be relaxed. However, the
flat mass-radius profile extends to greater distances as well,
implying that the final orbital separation will also be increased, so
that less orbital energy may actually be available for ejecting the gas.
Thus, successful envelope ejection for more evolved configurations is
not as favorable as might be expected.

In all of our simulations the mass ejection is concentrated toward the
equatorial plane with about five times as much mass loss in the
orbital plane as compared to the polar direction.  In addition, the
results indicate that an expanding disk can form briefly about the remnant
binary system.  These morphologies may be consistent with the
equatorial-to-polar density contrast required by the interacting winds
model for the shaping of planetary nebulae.  Specifically, the results
described in Frank et al. (1993) and Mellema \& Frank (1995) indicate
that variations of three to five are adequate in providing an envelope
morphology which resembles those planetary nebula systems exhibiting a
clear axisymmetric structure.  Furthermore, the presence of a
geometrically thick disk may facilitate the formation of jets in
planetary nebulae (Soker \& Livio 1994).

The results obtained from these numerical simulations can be used to
place constraints on the progenitor systems of planetary nebulae with
binary nuclei, provided that orbital angular momentum losses from the
remnant binary are insignificant following the common envelope phase.
Angular momentum losses associated with either magnetic braking or
gravitational radiation are unimportant for the binary stars within
planetary nebulae since the age of the nebula (and hence the age of
the post-common-envelope system) is characteristically less than about
$10^4$ years.  Orbital evolution due to angular momentum transfer from
the orbit to a circumbinary disk is also not expected to be
significant since our results demonstrate that the majority of mass in
the disk is situated at great distances (many times the orbital
separation) from the binary system.  Furthermore, the angular momentum
loss from the orbit responsible for the spin-up and ejection of the
remaining matter in the vicinity of the remnant binary ($\lapprox 0.01
\msun$) can be estimated from the study of Shu, Lubow, \& Anderson
(1979), and it is expected to reduce the orbital separation only slightly
(by about 10\%).

\begin{figure}[t]
\hspace*{1.3in}
\epsfxsize=9 true cm
\epsfysize=9 true cm
\epsffile{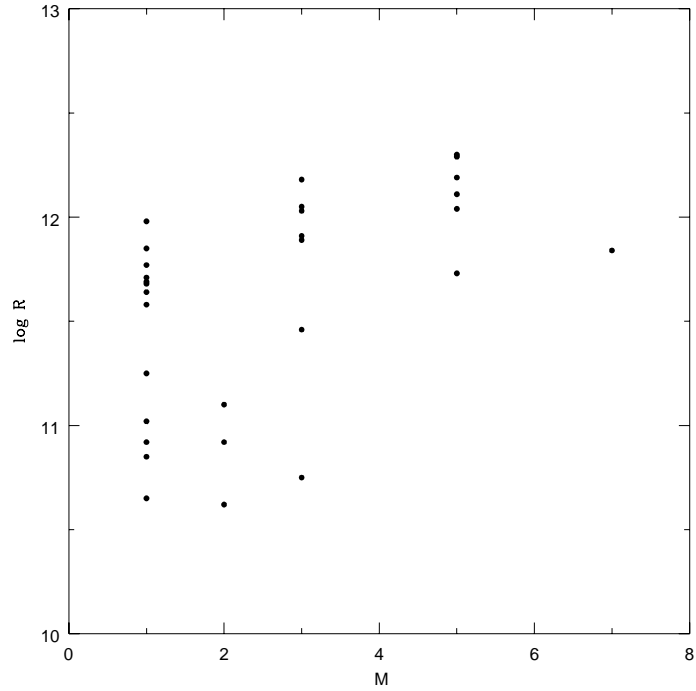}
\figcaption{Orbital separation of a post-common-envelope binary system 
as a function of the red giant mass in solar units.  The separation is
assumed to be a factor of 6 less than the radius of the common envelope
for which the logarithmic derivative of the pressure with respect to 
the radius is a minimum. 
\label{taamfig}}
\end{figure}

The observational study of binary stars within planetary nebulae
reveals that many systems (6 out of the 12 known, although this is
affected by selection effects) have orbital periods less than 0.7 days
(see Bond 1995).  A total mass for the remnant binary of 1 $\msun$
implies that the orbital separation following the mass ejection phase
of the common envelope is $3.3 \rsun$.  Based on the present study and
following the work of Terman \& Taam (1996) on the estimated final
orbital separations of post-common-envelope systems, it is found that
white dwarf masses $< 0.6 \msun$ are indicated for systems with
orbital periods less than 0.7 days.  (For systems with orbital periods
$\gtrsim 0.7$ days, no such constraint can be placed on the white
dwarf masses).  It should be noted that white dwarf masses in the
range from about 0.46 - 0.56 $\msun$ cannot be formed in the common
envelope scenario since such core masses develop during the core
helium burning phase of the red giant when the red giant is
insufficiently large in radius to initiate mass transfer (see Webbink
1976). The progenitor masses leading to systems with orbital periods
$\lapprox 0.7$ days are less than $3 \msun$ as illustrated in
Figure~\ref{taamfig}.  Based on the present work, the final orbital
separations have been estimated by assuming that they are a factor of
6 less than the radial distance in the giant corresponding to the
point where the logarithmic derivative of the pressure with respect to
radius is a minimum (this point marks the outer boundary of the flat
portion of the flat mass-radius profile; see Terman \& Taam 1996).
Greater orbital separations correspond to progenitor stars in more
advanced evolutionary phases with more massive cores.  The result that
lower white dwarf masses result from the lower mass range of
intermediate mass stars is consistent with the seminal work of
Paczynski (1970).  The radii of these progenitor stars are $\sim
10^{13}$ cm corresponding to orbital periods of the progenitor systems
at the onset of the common envelope phase of $\lapprox 0.5$ years.

To further our understanding of the outcome of the common envelope
phase, higher resolution and longer term evolutionary studies will be
necessary to follow the total ejection of the common envelope.
Systematic studies will be required to determine the efficiency of
mass ejection and to delineate the parameter space separating those
systems which merge from those systems which survive the common
envelope stage.  Such studies will also illustrate the diversity of
morphologies of ejected matter (formation of disks and density
contrasts between the polar and equatorial directions) relevant to the
shaping of planetary nebulae.

\acknowledgements

This work has been supported by NSF grant AST-9415423.

\clearpage
\begin{table*}
\begin{center}
\newdimen\digitwidth
\setbox0=\hbox{\rm 0}
\digitwidth=\wd0
\catcode`?=\active
\def?{\kern\digitwidth}
\begin{tabular}{ccccccc}
\tableline
\tableline
Sequence & $M_1(\msun)$ & $M_c(\msun)$ & $M_2(\msun)$ &  $f_{sync}$  \\
\tableline
    1    &     3      &  0.7  &      0.4     &     1.0   \\
    2    &     3      &  0.7  &      0.4     &     0.0   \\
    3    &     5      &  1.0  &      0.4     &     1.0   \\
    4    &     5      &  1.0  &      0.6     &     1.0   \\
    5    &     5      &  0.94 &      0.6     &     0.0   \\
\end{tabular}
\end{center}
\caption{Initial Parameters of Common Envelope Sequences}
\label{1}
\end{table*}
\bigskip

\begin{table*}
\begin{center}
\newdimen\digitwidth
\setbox0=\hbox{\rm 0}
\digitwidth=\wd0
\catcode`?=\active
\def?{\kern\digitwidth}
\begin{tabular}{ccccccc}
\tableline
\tableline

Sequence & $a (\rsun)$ & $P$ (days) &  $v$ (km s$^{-1}$)  & $\alpha_{CE}$\\
\tableline
    1    &   4.42      &    1.03    &     3      &   0.38 \\
    2    &   4.68      &    1.12    &     4      &   0.46 \\
    3    &   4.37      &    0.90    &     4      &   0.24 \\
    4    &   4.78      &    0.96    &     1      &   0.44 \\
    5    &   8.91      &    2.48    &     2      &   0.53 \\
\end{tabular}
\end{center}
\caption{Final Parameters of Common Envelope Sequences}
\label{2}
\end{table*}

\clearpage

\begin{references}

\reference{} Berger, M. J. \& Colella, P. 1989, J. Comp. Phys., 82, 64

\reference{} Berger, M. J. \& Oliger, J. 1984, J. Comp. Phys., 53, 484

\reference{} Bodenheimer, P. \& Taam, R. E. 1984, \apj, 280, 771

\reference{} Bond, H. E. 1995, in Asymmetrical Planetary Nebulae, ed. A. 
Harpaz \& N. Soker (Bristol: Institute of Physics Publishing), 61

\reference{} Bond, H. E. \& Livio, M. 1990, \apj, 355, 568

\reference{} Burkert, A. \& Bodenheimer, P. 1993, \mnras, 264, 798

\reference{} Counselman, C. C. 1973, \apj, 180, 307

\reference{} Darwin, G. H. 1879, Proc. R. Soc. London, 29, 168

\reference{} de Kool, M. 1987, Ph.D. Thesis, University of Amsterdam

\reference{} de Kool, M. 1992, \aap, 261, 188

\reference{} de Kool, M. 1996, in Evolutionary Processes in Binary Stars,
ed. R. A. M. J. Wijers, M. B. Davies, \& C. A. Tout, (Dordrecht:
Kluwer), 365

\reference{} Delgado, A. J. 1980, \aap, 87, 343

\reference{} Eggleton, P. P. 1971, \mnras, 151, 351

\reference{} Eggleton, P. P. 1972, \mnras, 156, 361

\reference{} Frank, A., Balick, B., Icke, V., \& Mellema, G. 1993, \apj, 404, L25

\reference{} Gingold, R. A. \& Monaghan, J. J. 1977, \mnras, 181, 375

\reference{} Han, Z., Podsiadlowski, P., \& Eggleton, P. P. 1995, \mnras, 272, 800

\reference{} Hjellming, M. S. \& Taam, R. E. 1991, \apj, 370, 709

\reference{} Iben, I. \& Livio, M. 1993, \pasp, 105, 1373

\reference{} Iben, I. \& Tutukov, A. V. 1984, \apjs, 54, 335

\reference{} Kalogera, V. \& Webbink, R. F. 1996, \apj, 458, 301

\reference{} Kley, W., Shankar, A., \& Burkert, A. 1995, \aap, 297,
739

\reference{} Kopal, Z. 1978, Dynamics of Close Binary Systems,
(Dordrecht: Reidel)

\reference{} Kwok, S. 1982, \apj, 258, 280

\reference{} Lai, D., Rasio, F. A., \& Shapiro, S. L. 1993, \apj, 406, L63

\reference{} Lai, D., Rasio, F. A., \& Shapiro, S. L. 1994, \apj, 420, 811

\reference{} Livio, M. 1995, in Asymmetrical Planetary Nebulae, ed. A. 
Harpaz \& N. Soker (Bristol: Institute of Physics Publishing), 51

\reference{} Livio, M. \& Soker, N. 1984, \mnras, 208, 763.

\reference{} Livio, M. \& Soker, N. 1988, \apj, 329, 764

\reference{} Lucy, L. 1977, \aj, 82, 1013

\reference{} Mellema, G. \& Frank, A. 1995, in
Asymmetrical Planetary Nebulae, ed. A. Harpaz \&
N. Soker (Bristol: Institute of Physics Publishing), 229

\reference{} Meyer, F. \& Meyer-Hofmeister, E. 1979, \aap, 78, 179

\reference{} Monaghan, J. J. 1985, Computer Physics Reports, 3, 71

\reference{} Monaghan, J. J. 1992, \araa, 30, 543

\reference{} Paczynski, B. 1970, Acta Astr., 20, 47

\reference{} Paczynski, B. 1976, in IAU Symposium No. 73, The Structure and 
Evolution of Close Binary Systems, ed. P. Eggleton, S. Mitton, \& J. Whelan,
(Dordrecht: Reidel), 75

\reference{} Paczynski, B. \& Sienkiewicz, R. 1972, Acta Astr., 22, 73

\reference{} Rasio, F. \& Livio, M. 1996, \apj, 471, 366

\reference{} Ruffert, M. 1993, \aap, 280, 141

\reference{} Shu, F. H., Lubow, S. H., \& Anderson, L. 1979, \apj, 229, 223

\reference{} Sod, G. A. 1978, J. Comp. Phys., 27, 1

\reference{} Soker, N. \& Livio, M. 1989, \apj, 339, 268

\reference{} Soker, N. \& Livio, M. 1994, \apj, 421, 219

\reference{} Taam, R. E. \& Bodenheimer, P. 1989, \apj, 337, 849

\reference{} Taam, R. E. \& Bodenheimer, P. 1991, \apj, 373, 246

\reference{} Taam, R. E., Bodenheimer, P., \& Ostriker, J. P. 1978,
\apj, 222, 269

\reference{} Terman, J. L. \& Taam, R. E. 1996, \apj, 458, 692

\reference{} Terman, J. L, Taam, R. E., \& Hernquist, L. 1994, \apj, 422, 729

\reference{} Terman, J. L, Taam, R. E., \& Hernquist, L. 1995, \apj, 445, 367

\reference{} Tutukov, A. V. \& Yungelson, L. R. 1979, Acta Astr., 29, 666

\reference{} van den Heuvel, E. P. J. 1987, in High Energy
Phenomena Around Collapsed Stars, ed. F. Pacinin (Dordrecht: Reidel), 1

\reference{} van Leer, B. 1977, J. Comp. Phys., 23, 276

\reference{} von Neumann, J. \& Richtmeyer, R. D. 1950, J. Appl. Phys., 21, 232

\reference{} Webbink, R. F. 1976, in IAU Symposium No. 73, Structure and 
Evolution of Close Binary Systems, ed. P. Eggleton, S. Mitton, \& J. 
Whelan, (Dordrecht: Reidel), 207

\reference{} Webbink, R. F. 1979, in IAU Coll. No. 46, Changing
Trends in Variable Star Research, ed. F. M. Bateson, J. Smak, \& I.
H. Urch (Hamilton, NZ: Univ. Waikato), 102

\reference{} Yorke, H. W., Bodenheimer, P., \& Laughlin, G. 1993, \apj, 411, 274

\reference{} Yorke, H. W., Bodenheimer, P. \& Taam, R. E. 1995, \apj, 451, 308

\reference{} Yungelson, L. R., Tutukov, A. V., \& Livio, M. 1993, \apj, 418, 794
\end{references}
\end{document}